\documentclass[conference]{IEEEtran}
\usepackage{cite}
\usepackage[numbers]{natbib}
\usepackage{amsmath,amssymb,amsfonts}
\usepackage{algorithmic}
\usepackage{graphicx}
\usepackage{flushend}
\usepackage{dblfloatfix}
\usepackage{textcomp}
\usepackage{xcolor}
\usepackage{url}
\usepackage{tcolorbox}
\usepackage{subfigure}
\usepackage{multirow}
\usepackage{tikzsymbols}
\usepackage[colorinlistoftodos]{todonotes}
\usepackage{ulem}
\usepackage{balance}
\usepackage{colortbl}
\usepackage{booktabs}
\usepackage{makecell}
\usepackage{authblk}
\usepackage[T1]{fontenc}

\usepackage{scalerel,graphicx,xparse}

\NewDocumentCommand\emojiThumbsUp{}{\scalerel*{\includegraphics{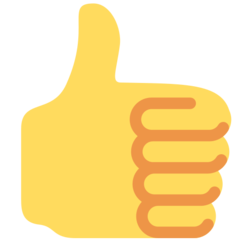}}{X}}
\NewDocumentCommand\emojiThumbsDown{}{\scalerel*{\includegraphics{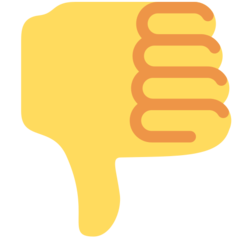}}{X}}
\NewDocumentCommand\emojiTada{}{\scalerel*{\includegraphics{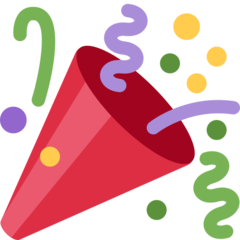}}{X}}
\NewDocumentCommand\emojiHeart{}{\scalerel*{\includegraphics{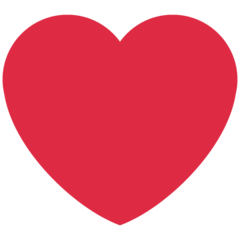}}{X}}
\NewDocumentCommand\emojiRocket{}{\scalerel*{\includegraphics{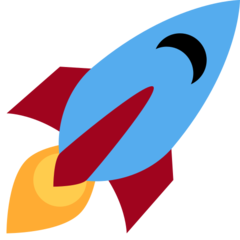}}{X}}
\NewDocumentCommand\emojiConfused{}{\scalerel*{\includegraphics{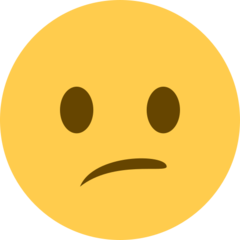}}{X}}
\NewDocumentCommand\emojiEyes{}{\scalerel*{\includegraphics{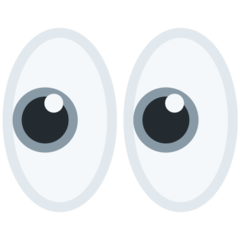}}{X}}
\NewDocumentCommand\emojiGrinning{}{\scalerel*{\includegraphics{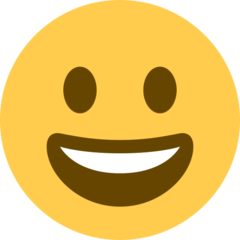}}{X}}

\newcommand\son[1]{{\textcolor{black}{#1}}}
\newcommand\thefont{\expandafter\string\the\font}
\def\BibTeX{{\rm B\kern-.05em{\sc i\kern-.025em b}\kern-.08em
    T\kern-.1667em\lower.7ex\hbox{E}\kern-.125emX}}

\newcommand\rev[2]{{\color{black}#2}}

\begin{document}

\definecolor{gray(x11gray)}{rgb}{0.75, 0.75, 0.75}
 	
\def\mybar#1{
  {\color{gray}\rule{#1cm}{8pt}}}

\title{More Than React:
Investigating The Role of Emoji Reaction in GitHub Pull Requests}

\author{
Teyon Son,
Tao Xiao,
Dong Wang, 
Raula Gaikovina Kula,
Takashi Ishio,
Kenichi Matsumoto\\
Nara Institute of Science and Technology, Japan \\
Email: \{son.teyon.sr7, tao.xiao.ts2, wang.dong.vt8, raula-k, ishio, matumoto\}@is.naist.jp}


\maketitle

\begin{abstract} 
Context: {Open source software development has become more social and collaborative, especially with the rise of social coding platforms like GitHub. 
Since 2016, GitHub started to support more informal methods such as emoji reactions, with the goal to reduce commenting noise when reviewing any code changes to a repository.
Interestingly, preliminary results indicate that emojis do not always reduce commenting noise (i.e., eight out of 20 emoji reactions), providing evidence that developers use emojis with ulterior intentions.
\son{From a reviewing context, the extent to which emoji reactions facilitate for a more efficient review process is unknown.}
}\\
Objective:
In this registered report, 
we introduce the study protocols to investigate ulterior intentions and usages of emoji reactions, apart from reducing commenting noise during the discussions in GitHub pull requests (PRs).
As part of the report, we first perform a preliminary analysis to whether emoji reactions can reduce commenting noise in PRs and then introduce the execution plan for the study.\\
Method: We will use a mixed-methods approach in this study, i.e., quantitative and qualitative, with three hypotheses to test.
\end{abstract}

\section{Introduction}
In the past few years, open source software development has become more social and collaborative.
\son{Known as social coding,open source development promotes formal and informal collaboration by}
empowering the exchange of knowledge between developers~\cite{dabbish2012social}. 
GitHub, one of the most popular social coding platforms, attracts more than 72 million developers collaborating across 233 million repositories.\footnote{\url{https://github.com/search}}
Every day, thousands of people engage in conversations about code, design, bugs, and new ideas on GitHub. 
To promote collaboration, GitHub implements a vast number of social features (i.e., follow, fork, and stars).

Since 2016, GitHub introduced a new social function called ``\textit{reaction}'' for developers to quickly express their feelings in issue reports and PRs.
Especially for discussing a \son{Pull Requests(PRs)}, we find that\footnote{\url{https://tinyurl.com/3rpdr6dp}}:
\begin{quote}
    \textit{``In many cases, especially on popular projects, the result is a long thread full of emoji and not much content, which makes it difficult to have a discussion. With reactions, you can now reduce the noise in these threads'' - GitHub}
\end{quote}

In the context of code review, we assume that a thread full of emoji may also contribute to the existing forms of confusion for reviewers during the code review process.
For instance, \citet{confusion_saner_2019} pointed out that confusion delays the merge decision decreases review quality, and results in additional discussions.
\citet{hirao2020code} found that patches can receive both positive and negative scores due to the disagreement between reviewers, which leads to conflicts in the review process.

\begin{figure*}[t]
    \centering
    \subfigure[Example of emoji reaction reduce commenting noise.\label{fig:em}]{\includegraphics[width=0.65\textwidth]{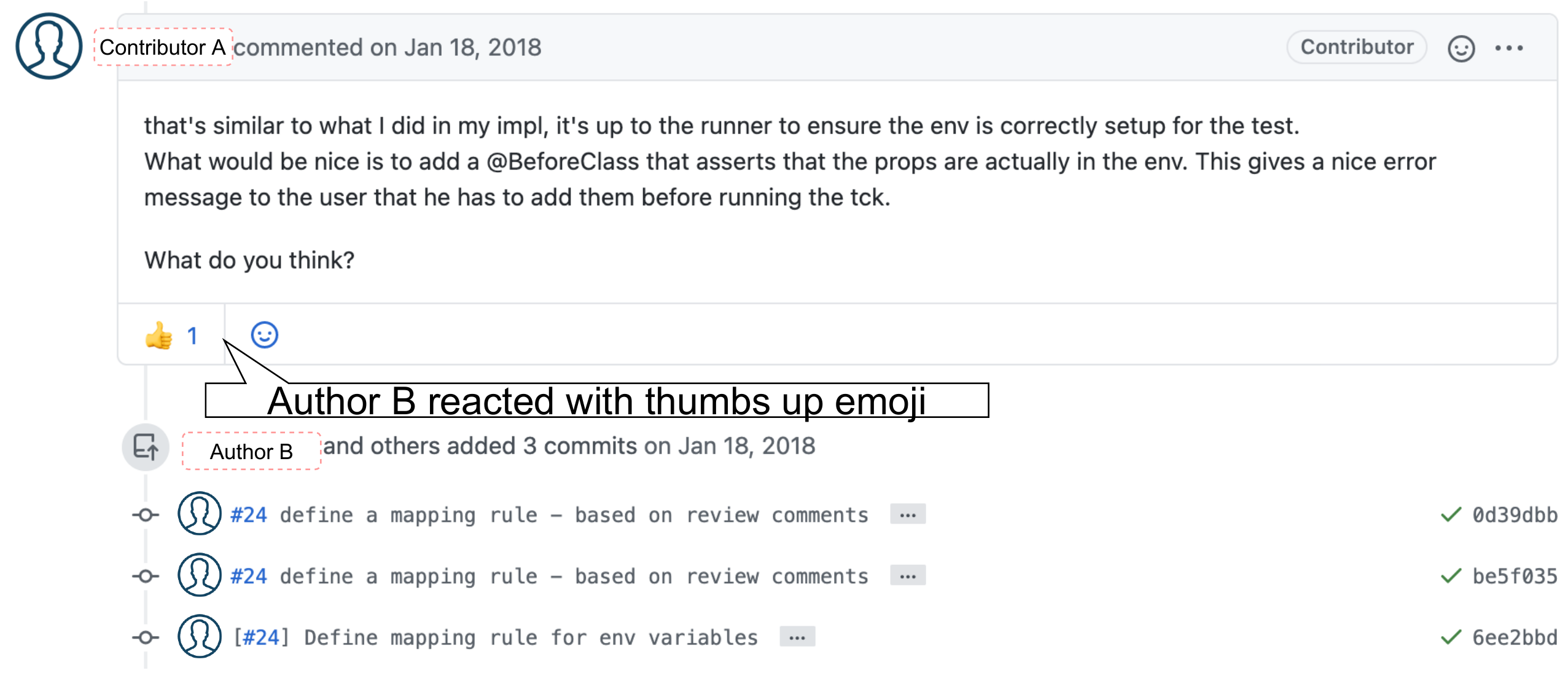}}
        \subfigure[Example of emoji reaction does not reduce commenting noise.\label{fig:moti}]{\includegraphics[width=0.65\textwidth]{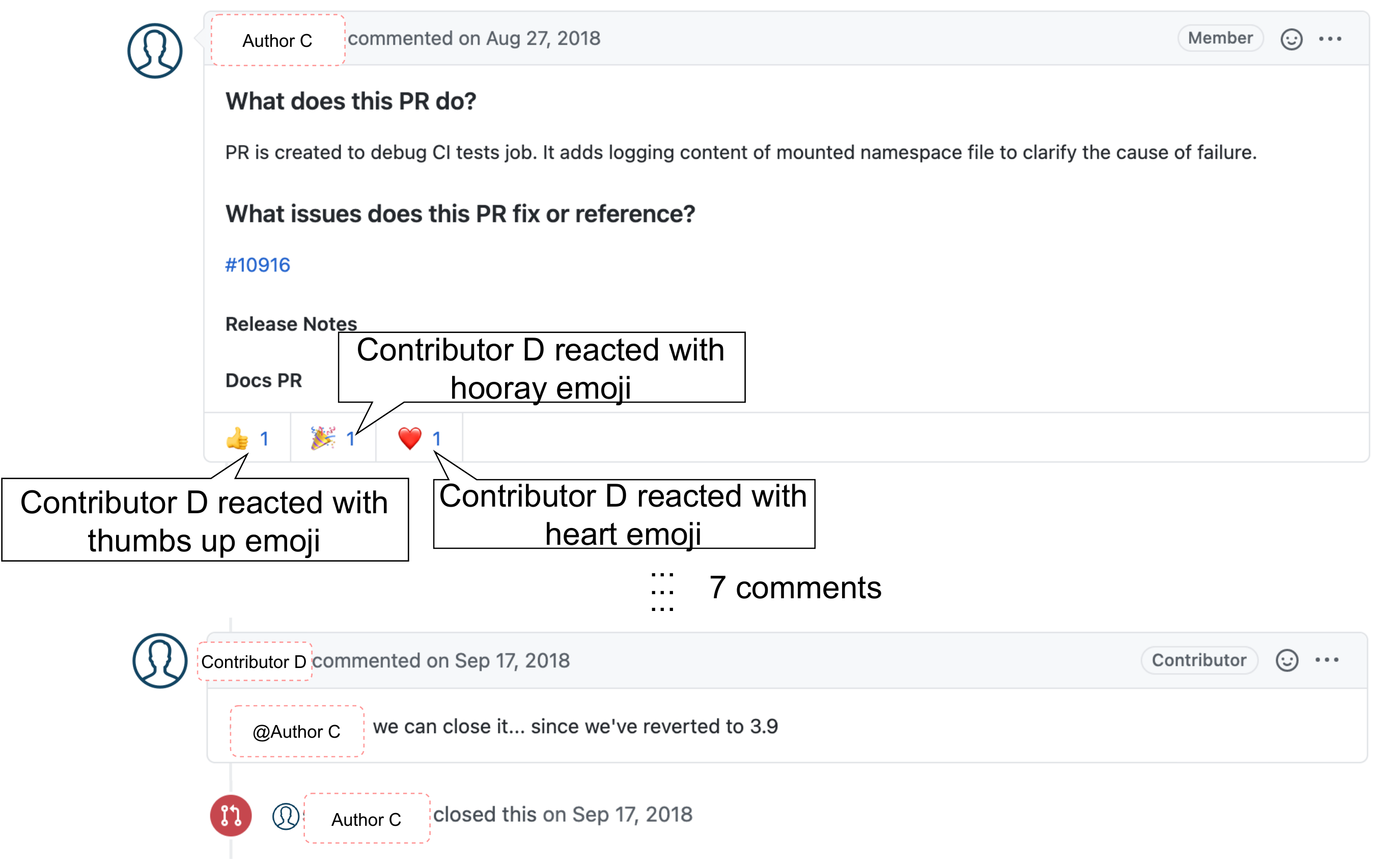}}
    \caption{ Examples of emoji reactions used in GitHub.}
    \label{fig:examples}
\end{figure*}

The Figure \ref{fig:examples} depicts two typical cases where the emoji reactions occur.
Figure \ref{fig:em} shows the case where the reaction does reduce unnecessary commenting in the thread, hence may lead to less confusion and conflicts.
The example illustrates how \texttt{Author B} reduces the commenting by simply reacting with a quick expression of approval through \texttt{THUMBS UP} \emojiThumbsUp.
In contrast, as shown in Figure~\ref{fig:moti}, there exists a case where the emoji usage has an ulterior intention and does not reduce comments in the discussion thread. 
In detail, \texttt{Contributor D} uses three positive emoji reactions (\texttt{THUMBS UP} \emojiThumbsUp, \texttt{HOORAY} \emojiTada, and \texttt{HEART} \emojiHeart) to represent the appreciation to this PR.
Then later goes on to provide detailed comments on the PR. 
We posit that the intention of the emoji was to express appreciation for the PR, and did not reduce the amount of commenting in the thread discussions.
As part of our preliminary study, we also found other cases where the emoji did not always reduce the commenting in the discussion. 
Under a closer manual inspection of 20 emoji reactions, we find that there are eight cases where the emoji reactions did not reduce commenting noise.

Therefore, \son{in this registered report}, we present our study protocol to investigate ulterior intentions and usages of emoji reactions, apart from reducing commenting noise during the discussions.
Specifically, we would like to (i) investigate the effect of emoji reaction related factors on the pull request process (i.e., review time), (ii) investigate whether the first time pull request is more likely to receive reactions, (iii) analyze the relationship between the reaction and the intention of comment, and \rev{R1.1}{(iv) explore consistency between sentiments of an emoji reaction and the sentiment of the comment.}
To enable other researchers to extend our study, we plan to make the study data publicly available.

\section{Preliminary Study}

The goal of our preliminary study is to explore \son{the extent to which emoji reaction usage reduces commenting noise of a PR}.
We selected the Eclipse projects as our case study since it is a mature thriving open source project with a number of contributors that actively submit and merge PRs.

\begin{table}[t]
  \caption{Preliminary Dataset Summary Statistics}
  \label{table:REPOSITORY AND PULL REQUEST IN ECLIPSE}
  \centering
  \resizebox{.5\textwidth}{!}{
  \begin{tabular}{lr@{}rr@{}rr@{}r}
    \toprule
    & \multicolumn{2}{c}{\# Repositories} &   \multicolumn{2}{c}{\# PR}  &  \multicolumn{2}{c}{\# PR Comments} \\
    \midrule
    With reactions & 203 & (48\%) & 6,867 & (8\%) & 9,256 & (4\%)   \\

    Without reactions & 217 & (52\%)  & 76,202 & (92\%)  & 249,354 & (96\%) \\

    Total& 420 & (100\%)  & 83,069 & (100\%) & 258,610 & (100\%)\\
    \bottomrule
  \end{tabular}}
\end{table}

\subsection {Data Collection}
We collected a list of 683 Eclipse repositories by using the official API of GitHub~\cite{restapi}. 
Since we focus on the reactions that are used in PRs,we excluded the repositories that do not have any PRs.
We obtained 420 repositories that have 83,069 PRs.
Then, we extracted PRs with reactions using GitHub GraphQL API~\cite{graphql}, along with the information of reaction types, reaction time, the developer who posts the reaction.
We excluded those reactions or PRs that are posted by bots since we investigate the reactions used by developers.
Based on our manual check, we \son{performed} the following two exclusions:
(i) \son{exclude developer name by using a regular expression matching(`github.app')}
(ii) \son{exclude the dependabot, a popular bot used to automatically notify developers of dependency upgrades.\footnote{\url{https://dependabot.com/}}}
In the end, we \son{obtained} 9,256 comments having emoji reactions across 203 Eclipse repositories, as shown in Table \ref{table:REPOSITORY AND PULL REQUEST IN ECLIPSE}.
\subsection{Data Analysis}

\begin{figure}[t]
    \centering
    \includegraphics[width=\linewidth]{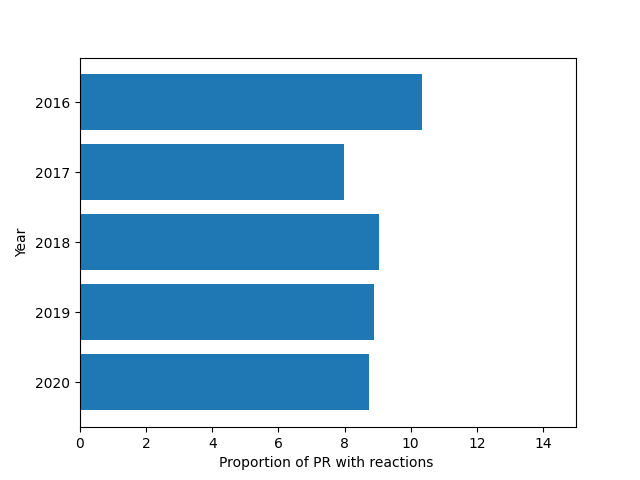}
    \caption{Proportion of PRs with reactions by year in Eclipse repositories.}
    \label{fig:prev_of_emoji}
\end{figure}

With our preliminary dataset, we conducted three exploratory analyses related to emoji reaction usage.
First, 
we investigate the prevalence of reaction usage yearly since this feature was initially introduced in 2016.
To do so, we measured the proportion of PRs that have at least one emoji reaction.
Second, 
we investigate what are the common reactions for developers to express during the PR process.
To do so, we grouped the eight existing emoji reactions into four categories:
\begin{itemize}
    \item \texttt{Positive} - is the single usage or the combination usage of \texttt{THUMBS UP} \emojiThumbsUp , \texttt{LAUGH} \emojiGrinning, \texttt{HOORAY} \emojiTada, \texttt{HEART} \emojiHeart, and \texttt{ROCKET} \emojiRocket reactions.
    \item \texttt{Negative} - is the single usage or the combination usage of \texttt{THUMBS DOWN} \emojiThumbsDown \,and \texttt{CONFUSED} \emojiConfused. 
    \item \texttt{Neutral} - is the usage of \texttt{EYES} \emojiEyes reaction.
    \item \texttt{Mixed} - is the combination usage of the four categories mentioned above.
\end{itemize}
For example, if a PR comment is reacted by the \texttt{THUMBS UP}  \emojiThumbsUp \, and the \texttt{THUMBS DOWN} \emojiThumbsDown, we then classify this case as \textit{Mixed}. 
Third, 
we further investigate whether or not the reaction is used to reduce commenting noise during the review process.
To do so, we randomly select 20 PR comment samples from the preliminary dataset and did a manual classification (i.e., reduce commenting noise or not) among the first four authors.
After the emoji reactions were posted, if there are no additional comments related to the existing topic by the developers who react, we classify such case as \textit{Reduce Noise}.
Otherwise, we classify the case as \textit{Not Reduce Noise}.
For example, in Figure~\ref{fig:em}, the \texttt{Author B} reacted with \texttt{THUMBS UP} to the suggestion provided by the \texttt{Contributor A} and added three commits without any additional comments.
This case is labeled as \textit{Reduce Noise}.

\textbf{Positive emoji reactions are widely used in PRs.}
Two preliminary results are summarized. First, we find that around 8\% to 10\% of PRs have at least one reaction in Eclipse repositories between 2016 and 2020.
Figure~\ref{fig:prev_of_emoji} shows the proportion of PRs with reactions yearly.
Second, we observe that most of the reactions in PR comments are \textit{Positive} (i.e., the single or combination usage of \emojiThumbsUp, \emojiGrinning, \emojiTada, \emojiHeart, and \emojiRocket), accounting for 98.1\%.
Upon a further inspection, among these \textit{Positive} reactions, the single usage of \emojiThumbsUp \, almost reached 86.5\%.
On the other hand, the usage of \textit{Negative}, \textit{Neutral}, and \textit{Mixed} only accounts for 1.84\%.
Table~\ref{distribtuion_emoji} shows the distribution of the sentiment of emoji reaction usage, indicating that the positive reaction is the most prevalent.

\textbf{Emoji reactions do not always reduce the commenting noise.}
Table~\ref{table:prevalence} shows the frequency of samples where whether reactions reduce commenting noise from our manual classification.
The table shows that eight samples (40\%) are classified as \textit{Not Reduce Noise}.
In these samples, after the emoji reactions were posted, developers post additional comments to express and discuss issues on the PR. 

\begin{tcolorbox}
\textbf{Summary:} 
Preliminary results show that around 8\% to 10\% of PRs have reactions in Eclipse repositories. 
We find cases where the emoji did not always reduce commenting noise in the discussion. 
Under a closer manual inspection of 20 emoji reactions, we find that there are eight cases where the emoji reactions did not reduce commenting noise. 
\end{tcolorbox}


\begin{table}[t]
\centering
\caption{Distribution of Sentiments of Emoji Reactions.}
\label{distribtuion_emoji}
\begin{tabular}{lr@{}rl}
\toprule
Emoji reaction sentiments & \multicolumn{2}{c}{\# PR Comments} & \\
\midrule
Positive                   & 9,084                 & (98.10\%) & \mybar{0.98}                      \\

Negative                   & 67                    & (0.72\%)  & \mybar{0.0072}                       \\

Neutral                    & 74                    & (0.79\%) &  \mybar{0.0079}                      \\

Mixed                & 31                    & (0.33\%)      &  \mybar{0.0033}                  \\
\midrule
Total                      & 9,256                 & (100\%)               \\   
\bottomrule
\end{tabular}
\end{table}

\begin{table}[]
	\fontsize{9}{11}\selectfont
	\tabcolsep=0.2cm
\centering
\caption{The frequency count of manual samples where whether the emoji reactions reduced the commenting noise.}
    \label{table:prevalence}
    \vspace{-.2cm}
\begin{tabular}{lrll}
\toprule
\textbf{Category}                            & \textbf{Count} &  \\ \midrule
Reducing commenting noise    &   12 & \mybar{0.6}   \\
Not Reducing commenting noise  &    8&\mybar{0.4}  \\\bottomrule
\end{tabular}
\vspace{-.4cm}
\end{table}

\section{Study Protocols}
\label{AA}
In this section, we present the design of our study. This section consists of our research questions with their motivations.

\subsection{Research Questions}
Inspired by the motivating examples and the preliminary study, we formulate four research questions to guide our study:
\begin{itemize}
\item \textbf{RQ1: \textit{Does the emoji reaction used in the review discussion correlate with review time?}}\\ 
Prior studies \cite{Olga_2016, Shopify_2018} have widely analyzed the impact of technical and non-technical factors on the review process (e.g., review outcome, review time). 
However, little is known about whether or not the emoji reaction can be correlated with review time.
It is possible that emoji reaction may shorten the review time, as it could reduce the noise during the review discussions.
Thus, our motivation for the first research question is to explore the correlation between the emoji reaction used in the review discussion and review time.

\item\textbf{RQ2: \textit{Does a PR submitted by a first-time contributor receive more emoji reactions?}}\\ 
As shown in Figure \ref{fig:moti}, we find that the emoji reaction might be used to express appreciation for submitting a PR.
Our motivation for this research question is to understand if contributors that have never submitted to the project before receive more emoji reactions.
Furthermore, answering this research question will provide insights into a \son{potent} ulterior motive for \son{a emoji reaction.}

Our assumption is that:
\begin{quote}
  \textit{ H1: PRs submitted by first-time contributor receive more emoji reactions.} 
  Existing contributors express positive feelings to attract newcomers to the project.
\end{quote}

\item\textbf{RQ3: \textit{What is the relationship between the intention of comments and their emoji reaction?}}\\ 
Our preliminary study findings show that emoji reactions do not always reduce the commenting noise.
Hence, our motivation for the third research question is to explore the relationship between the intention of comments and their reactions. 

Our assumption is that:
\begin{quote}
    \textit{H2: Most emojis are uniformly distributed across the different intentions.}
    Specific intentions may explain the ulterior purpose of reacting with an emoji reaction.
\end{quote}
\item\textbf{{RQ4: \textit{Is emoji reaction consistent with comment sentiment?}}}

\rev{R1.1,R1.4}{We found that specific sentiments of the emoji (i.e., \texttt{THUMBS UP} \emojiThumbsUp) are widely used in PRs from our preliminary study. Our motivation for this research question is to investigate whether there is any inconsistency between sentiments of the comments and sentiments of the emoji reactions. Furthermore, we plan to manually check the reasons why inconsistency happened. We believe answering RQ4 would help newcomers better understand the emoji usage in the PR discussion.}

Our assumption is that: 
\begin{quote}
    \textit{H3: The sentiment of emoji reactions are uniformly distributed across the same comment sentiments.}
    Specific sentiments may explain the ulterior purpose of reacting with an emoji. This may be useful in understanding what information is needed in code review. 
\end{quote}
\end{itemize}

\section{Data collection}
To generalize the results of the study, \rev{R2.2,R2.7,R3.1}{we plan to expand on our dataset from active software development repositories shared by Hata et al. \cite{hata}. Each repository in this dataset has more than 500 commits and at least 100 commits during the most active two-year period. In total, this dataset contains 25,925 repositories from seven languages (i.e.,  C, C++, Java, JavaScript, Python, PHP, and Ruby). We will use the GraphQL API~\cite{graphql} to obtain PRs created before March 13rd 2016 where GraphQL was introduced. The whole dataset will be used for all four research questions.}

\section{Execution Plan}
In this section, we present the execution plan of our experiment. 
We will use a mixed method consisting of both quantitative and qualitative analysis to answer our research questions.

\subsection{Research Method for RQ1:}

For the first research question, we plan to use a quantitative method. 
To investigate the effect of emoji reaction related factors on the pull request process (i.e., review time), we plan to perform a statistical analysis using a non-linear regression model.
This model allows us to capture the relationship between the independent variable and the dependent variable.
The goal of our statistical analysis is not to predict the review time but to understand the associations between the emoji reaction and the review time.

For the independent variables, similar to the prior studies~\cite{Shopify_2018, WangEMSE2021}, we will select the following confounding factors as our independent variables:
\begin{itemize}
    \item \textit{PR size:} The total numbers of added and deleted lines of code changed by a PR.
    \item \textit{Change file size:} The number of files what were changed by a PR.
    \item \textit{Purpose:} The purpose of a PR, i.e., bug, document, feature.
    \item \textit{\# Comments}: The total number of comments in a PR discussion thread.
    \item \textit{\# Author Comments}: The total number of comments by the author in a PR discussion thread.
    \item \textit{\# Reviewer Comments}: The total number of comments by the reviewers in a PR discussion thread.
    \item \textit{Patch author experience}: The number of prior PRs that were submitted by the PR author.
    \item \textit{Reviewers:} The number of developers who posted a comment to a review discussion.
    \item \textit{Commit size:} The number of commits in a PR.
\end{itemize}
Since we investigate the effect of the emoji reaction, we plan to compute additional independent variables that are related to emoji reaction:
\begin{itemize}
    \item \textit{With emoji reaction:} Whether or not a PR includes any emoji reaction (binary).
    \item \textit{The number of emoji reactions:} The count of emoji reaction in a PR.
\end{itemize}
For the dependent variable (i.e., review time), we measure the time interval in hours from the time when the first comment was posted until the time when the last comment was posted.
For the model construction, we will adopt the steps that are similar to the prior studies, including (i) Estimating budget for degrees of freedom, (ii) Normality adjustment, (iii) Correlation and redundancy analysis, (iv) Allocating degrees of freedom, and (v) Fitting statistical models.

\textit{\textbf{(a) Analysis Plan:}} 
\rev{R2.3}{We will analyze the constructed regression models in the following three steps:
(i) Assessing model stability. To evaluate the performance of our models, we will report the adjusted $R^2$~\cite{hastie2009elements}. We will also use the bootstrap validation approach to estimate the optimism of the adjusted $R^2$. (ii) Estimating the power of explanatory variables. Similar to prior work \cite{WangEMSE2021}, we plan to test the significant correlation of independent variables with p-value and employ Wald statistics to measure the impact of each independent variable.
(iii) Examining relationship. Finally, we will examine and plot the direction of the relationship between each independent variable (i.e., especially emoji reaction related variables) and the dependent variable.}


\subsection{Research Method for RQ2:}
For RQ2, we plan to use a quantitative method.
\rev{R1.3,R2.5}{
To do so, we will construct two groups of pull requests to compare against: first-time contributors and non-first time contributors (control group).
For the first-time contributor group, we will identify all pull requests that are submitted by first-time contributors from our dataset. 
For the non-first time contributor group, to construct a balanced control group, we will randomly select the equal number of pull requests that are submitted by non-first time contributors.
We will then divide the pull requests into ones having emoji reactions and the other ones without emoji reactions, respectively.
}

\textit{\textbf{(a) Analysis Plan:}} 
We  will  present  a  pivot  chart to show the frequency of pull requests having emoji reactions or without emoji reactions by first-time contributors and non-first time contributors.
The plot x-axis will represent two groups of first-time contributors and non-first time contributors.
Furthermore, each group will be divided into two parts: pull requests with emoji reaction and pull reactions without emoji reactions.
The plot y-axis will represent the frequency count of pull requests.

\textit{\textbf{(b) Significant Testing:}} \rev{R2.4}{To select a suitable statistical test, we will adopt the Shapiro-Wilk test with alpha = 0.05.
In the case when the p-value is greater than 0.05, we will perform a two-tailed independent t-test with alpha 0.05. 
Otherwise, we will adopt a two-tailed Mann Whitney U test~\citep{Mann} with alpha = 0.05 to validate.
}

\rev{R2.2}{In addition, we will investigate the effect size. In case when the data is normally distributed, we will use Hedges g effect size \cite{53d3fd7fd0d8428f8376b5a209affcfb}.
Effect size is analyzed as follows: (1) $|d|$~\textless~0.2 as Negligible, (2) 0.2~$\le$~$|d|$~\textless 0.5 as Small, (3) 0.5~$\le$ $|d|$~\textless 0.8 as Medium, or (4) 0.8~$\leq$~$|d|$~as Large. If the data are not normally distributed, we will apply Cliff's delta (Romano et al, 2006) to measure effect size. 
Effect size is analyzed as follows: (1) $|\delta|$~\textless~0.147 as Negligible, (2) 0.147~$\le$~$|\delta|$~\textless 0.33 as Small, (3) 0.33~$\le$ $|\delta|$~\textless 0.474 as Medium, or (4) 0.474~$\leq$~$|\delta|$~as Large.
}




\subsection{Research Method for RQ3:}
For RQ3, we plan to use a quantitative  method to classify the intentions of the comments. 
\rev{R1.2}{
To categorize the intentions of the comments, we will use a taxonomy of intention proposed by Huang \cite{huang2018automating}.
They manually categorized  5,408 sentences from issue reports of four projects in GitHub to generalize the linguistic pattern for category identification.}

The taxonomy of intention category is described below:
\begin{itemize}
\item \textbf{Information Giving (IG)}: Share knowledge and experience with other people, or inform other people about new plans/updates (e.g., ``The typeahead from Bootstrap v2 was removed.'').
\item \textbf{Information Seeking (IS)}: Attempt to obtain information or help from other people (e.g., ``Are there any developers working on it?'').
\item \textbf{Feature Request (FR)}: Require to improve existing features or implement new features (e.g., ``Please add a titled panel component to Twitter Bootstrap.'').
\item \textbf{Solution Proposal (SP)}: Share possible solutions for discovered problems (e.g., ``I fixed this for UI Kit using the following CSS.'').
\item \textbf{Problem Discovery (PD)}: Report bugs, or describe unexpected behaviors (e.g., ``the firstletter issue was causing a crash.'').
\item \textbf{Aspect Evaluation (AE)}: Express opinions or evaluations on a specific aspect (e.g., ``I think BS3’s new theme looks good, it’s a little flat style.'').
\item \textbf{Meaningless (ML)}: Sentences with little meaning or importance (e.g., ``Thanks for the feedback!'').
\end{itemize}
\rev{R1.2}{
To facilitate the automation, they proposed a convolution neural network based classifier with high accuracy.
For RQ3, we will use this classifier to automatically label the intention of the comments.
To evaluate the robustness of this classifier in our dataset, we will first use the proposed classifier to automatically classify the intentions of the randomly sampled 30 comments. 
Then, we will manually check whether the labeled intentions of these 30 comments are correct or not.
The result of this sanity check will be presented as a percentage of the false positive, under 10\% being considerable.
}

\textit{\textbf{(a) Analysis Plan:}} 
To analyze the relationship between the intention of comments and their emoji reaction, we will use the association rule mining technique.
To show the diversity of different intentions from the classification, we will draw a histogram plot.
To show the results of the relationship, a table will be drawn with descriptive statistics,including the criteria support and confidence.

\textit{\textbf{(b) Significant Testing:}} 
To inspect whether or not the classified intentions of comments are normally distributed, we will adopt the Shapiro-Wilk test with alpha = 0.05, which is widely used for the normality test. 
In addition, to inspect whether the intentions of comments are significantly different, we will use Kruskal-Wallis non-parametric statistical test~\citep{Kruskal-Wallis}.

\subsection{Research Method for RQ4:}
For RQ4, we plan to use a qualitative and quantitative method.
\rev{R1.1, R1.4}{First of all, we will use a quantitative method to investigate whether there is any inconsistency between sentiments of emoji reaction and sentiments of the comments.
We determine the sentiments of the emoji based on the definition we discussed in section 2. Hence, Emoji sentiment can be categorized into the following types:  \texttt{Positive}, \texttt{Negative}, \texttt{Neutral}, and \texttt{Mixed}.
To extract the sentiment of the first responses, we plan to use SentiStrength-SE~\cite{Islam2018SentiStrengthSEED}, the state-of-the-art sentiment analysis tool for software engineering text. Similar to the tool we plan to use for RQ3, the input is reacted comments, and the output is the sentiment score of the given comment. The sentiment score varies from -5 (very Negative) to 5 (very Positive).
Based on the above definition, we consider it as inconsistent if the sentiments of the comment are different from the sentiments of the emoji. 

Then, we will conduct qualitative analysis to explore the possible reasons for inconsistency between sentiments of emoji and sentiments of the comments. To do so, we will apply the open coding approach~\cite{charmaz2014constructing} to classify the reasons for inconsistency. To discover as a complete list of reasons as possible, we strive for \textit{theoretical saturation}~\cite{eisenhardt1989building}.
Similar to prior work~\cite{hirao2019review},
we set our saturation criterion to 50, i.e., we continue to code randomly selected comments until no new reasons have been discovered for 50 consecutive comments. \rev{R2.6}{Furthermore, we perform the kappa agreement score~\cite{kappa} to evaluate the classification quality.
Similar to \citet{hata}, the agreement of the coding guide will be performed using a kappa agreement. Kappa result is interpreted as follows: values $\leq$ 0 as indicating no agreement and 0.01--0.20 as none to slight, 0.21--0.40 as fair, 0.41--0.60 as moderate, 0.61--0.80 as substantial, and 0.81--1.00 as almost perfect agreement.
The agreement scores larger than 0.81 (i.e., almost perfect) are considered for the manual analysis. Based on the prior experience, we estimate the size of the samples to range from 200-300 samples.}
}

\textit{\textbf{(a) Analysis Plan:}} 
\rev{R1.1,R1.4}{Similar to RQ3, we will depict a histogram plot to show the distribution of emoji types by sentiments of the comment.
To show the results of inconsistency reasons, we will draw a histogram plot to show the frequency. 
As part of our result presentation, we will paste the real examples during the analysis to describe the reason taxonomy.
}

\textit{\textbf{(b) Significant Testing:}} Similar to RQ3, we will use Shapiro-Wilk test to inspect whether or not the sentiments of emoji usage are normally distributed and use Kruskal-Wallis non-parametric statistical test to validate the significant difference.

\section{Implications}

We summarize our implications with the following take-away messages for the key stakeholders:
\begin{itemize}
\item \textbf{Researchers:} Answering RQ1 will help researchers understand the impact of emoji reactions, hence may contribute to existing knowledge on the code review process. 
As informal communication such as the usage of emojis becomes prevalent, it is a need to understand its role in keeping an efficient code review process.
We believe that emojis may also help remove toxic and other forms of anti-patterns \cite{anti} in the code review process. 
In terms of the intention of the emoji reactions, our study will complement all related works on emoji usage
\cite{chen2019sentimoji,chen2021emoji}.

\item \textbf{Contributors:} 
In terms of practitioners, we envision our study to assist projects to attract and maintain existing and potential contributors to the project. 
Especially for RQ2, the results of the study may provide some insights into how to attract newcomers and also how to provide a more friendly and welcoming environment.
Furthermore, answering RQ3 and RQ4 will provide some insights for contributors into undersanding the consistency of emoji reactions.
As emoji usage becomes popular~\cite{Li_2021}, the results of the study should provide guidelines on how to represent the common intentions when an emoji reaction is warranted.

\end{itemize}

\section{Threats to validity}
We identified three key threats to our study. 
First, the Eclipse projects that were used in our preliminary study may not be representative of all types of GitHub projects.
To increase generalizability, we will extend our study to include a sample of random GitHub projects \cite{hata}.
Our second threat is concerning the qualitative aspect of the study, as this is bias to human error in the classification.
This is because the interpretation of emoji usage may not be trivial.
To mitigate this, we employ the Kappa method to have multiple co-authors for agreement of each code. 
For instance, if positive emoji is used in an ironic context, it does not mean positive. This usage of emojis may influence our results.
Third, our quantitative analysis and data collection may include some false positives, such as bot reactions and comments. 
Currently, we manually exclude these bots for the preliminary study. To mitigate this, we plan to carefully identify and systematically remove bots based on official documentation.

\bibliographystyle{plainnat}
\bibliography{ownrefs}
\end{document}